\documentclass{article}
\usepackage{ismir,amsmath,cite,url}
\usepackage{graphicx}
\usepackage{color}
\usepackage{float}
\definecolor{ali_gray}{gray}{0.6}
\long\def\/*#1*/{}
\usepackage{subcaption}

\title{Conditioning Deep Generative Raw Audio Models for Structured Automatic Music}

\multauthor
{Rachel Manzelli$^*$ \hspace{1cm} Vijay Thakkar$^*$ \hspace{1cm} Ali Siahkamari \hspace{1cm} Brian Kulis} 
  {$^*$Equal contributions \hspace{1cm} ECE Department, Boston University\\
{\tt\small \{manzelli, thakkarv, siaa, bkulis\}@bu.edu}
}
\def\authorname{Rachel Manzelli, Vijay Thakkar, Ali Siahkamari, Brian Kulis}

\begin{document}

\maketitle
\begin{abstract}
\vspace{-.2cm}
Existing automatic music generation approaches that feature deep learning can be broadly classified into two types: raw audio models and symbolic models. Symbolic models, which train and generate at the note level, are currently the more prevalent approach; these models can capture long-range dependencies of melodic structure, but fail to grasp the nuances and richness of raw audio generations. Raw audio models, such as DeepMind's WaveNet, train directly on sampled audio waveforms, allowing them to produce realistic-sounding, albeit unstructured music. In this paper, we propose an automatic music generation methodology combining both of these approaches to create structured, realistic-sounding compositions. We consider a Long Short Term Memory network to learn the melodic structure of different styles of music, and then use the unique symbolic generations from this model as a conditioning input to a WaveNet-based raw audio generator, creating a model for automatic, novel music. We then evaluate this approach by showcasing results of this work.
\end{abstract}

\section{Introduction}\label{sec:introduction}
\vspace{-.2cm}
The ability of deep neural networks to generate novel musical content has recently become a popular area of research. Many variations of deep neural architectures have generated pop ballads,\footnote{http://www.flow-machines.com/} helped artists write melodies,\footnote{https://www.ampermusic.com/} and even have been integrated into commercial music generation tools.\footnote{https://www.jukedeck.com/} 

Current music generation methods are largely focused on generating music at the note level, resulting in outputs consisting of symbolic representations of music such as sequences of note numbers or MIDI-like streams of events. These methods, such as those based on Long Short Term Memory networks (LSTMs) and recurrent neural networks (RNNs), are effective at capturing medium-scale effects in music, can produce melodies with constraints such as mood and tempo, and feature fast generation times~\cite{biaxial, performance-rnn-2017}. In order to create sound, these methods often require an intermediate step of interpretation of the output by humans, where the symbolic representation transitions to an audio output in some way.

An alternative is to train on and produce raw audio waveforms directly by adapting speech synthesis models, resulting in a richer palette of potential musical outputs, albeit at a higher computational cost. WaveNet, a model developed at DeepMind primarily targeted towards speech applications, has been applied directly to music; the model is trained to predict the next sample of 8-bit audio (typically sampled at 16 kHz) given the previous samples~\cite{wavenet}. Initially, this was shown to produce rich, unique piano music when trained on raw piano samples. Follow-up work has developed faster generation times~\cite{fastwavenet}, generated synthetic vocals for music using WaveNet-based architectures~\cite{singing}, and has been used to generate novel sounds and instruments~\cite{nsynth}. This approach to music generation, while very new, shows tremendous potential for music generation tools. However, while WaveNet produces more realistic sounds, the model does not handle medium or long-range dependencies such as melody or global structure in music. The music is expressive and novel, yet sounds unpracticed in its lack of musical structure. 

Nonetheless, raw audio models show great potential for the future of automatic music. Despite the expressive nature of some advanced symbolic models, those methods require constraints such as mood and tempo to generate corresponding symbolic output~\cite{performance-rnn-2017}. While these constraints can be desirable in some cases, we express interest in generating structured raw audio directly due to the flexibility and versatility that raw audio provides; with no specification, these models are able to learn to generate expression and mood directly from the waveforms they are trained on. We believe that raw audio models are a step towards less guided, unsupervised music generation, since they are unconstrained in this way. With such tools for generating raw audio, one can imagine a number of new applications, such as the ability to edit existing raw audio in various ways.

Thus, we explore the combination of raw audio and symbolic approaches, opening the door to a host of new possibilities for music generation tools. In particular, we train a biaxial Long Short Term Memory network to create novel symbolic melodies, and then treat these melodies as an extra conditioning input to a WaveNet-based model. Consequently, the LSTM model allows us to represent long-range melodic structure in the music, while the WaveNet-based component interprets and expands upon the generated melodic structure in raw audio form. This serves to both eliminate the intermediate interpretation step of the symbolic representations and provide structure to the output of the raw audio model, while maintaining the aforementioned desirable properties of both models.

We first discuss the tuning of the original unconditioned WaveNet model to produce music of different instruments, styles, and genres. Once we have tuned this model appropriately, we then discuss our extension to the conditioned case, where we add a local conditioning technique to the raw audio model. This method is comparable to using a text-to-speech method within a speech synthesis model. We first generate audio from the conditioned raw audio model using well-known melodies (e.g., a C major scale and the Happy Birthday melody) after training on the MusicNet dataset~\cite{musicnet}.  We also discuss an application of our technique to editing existing raw audio music by changing some of the underlying notes and re-generating selections of audio. Then, we incorporate the LSTM generations as a unique symbolic component. We demonstrate results of training both the LSTM and our conditioned WaveNet-based model on corresponding training data, as well as showcase and evaluate generations of realistic raw audio melodies by using the output of the LSTM as a unique local conditioning time series to the WaveNet model.

This paper is an extension of an earlier work originally published as a workshop paper~\cite{manzelli_mume}. We augment that work-in-progress model in many aspects, including more concrete results, stronger evaluation, and new applications. 

\section{Background}
\vspace{-.2cm}
We elaborate on two prevalent deep learning models for music generation, namely raw audio models and symbolic models.

\subsection{Raw Audio Models}
Initial efforts to generate raw audio involved models used primarily for text generation, such as char-rnn\cite{char-rnn} and LSTMs. Raw audio generations from these networks are often noisy and unstructured; they are limited in their capacity to abstract higher level representations of raw audio, mainly due to problems with overfitting~\cite{gruv}.

In 2016, DeepMind introduced WaveNet~\cite{wavenet}, a generative model for general raw audio, designed mainly for speech applications. At a high level, WaveNet is a deep learning architecture that operates directly on a raw audio waveform.
In particular, for a waveform modeled by a vector $x = \{x_1, ... , x_T\}$ (representing speech, music, etc.), the joint probability of the entire
waveform is factorized as a product of conditional probabilities, namely
\begin{equation}\label{jointprob}
p({x}) = p(x_1)\prod_{t=2}^T p(x_t | x_1, ..., x_{t-1}).
\end{equation}
The waveforms in WaveNet are typically represented as 8-bit audio, meaning that each $x_i$ can take on one of 256 possible values. The WaveNet model uses a deep neural network to model the conditional probabilities $p(x_t|x_1,...,x_{t-1})$.  The model is trained by predicting values of the waveform at step $t$ and comparing them to the true value $x_t$, using
cross-entropy as a loss function; thus, the problem simply becomes a multi-class classification problem (with 256 classes) for each timestep in the waveform.

\begin{figure}[t]
\centering
\includegraphics[width=1 \columnwidth]{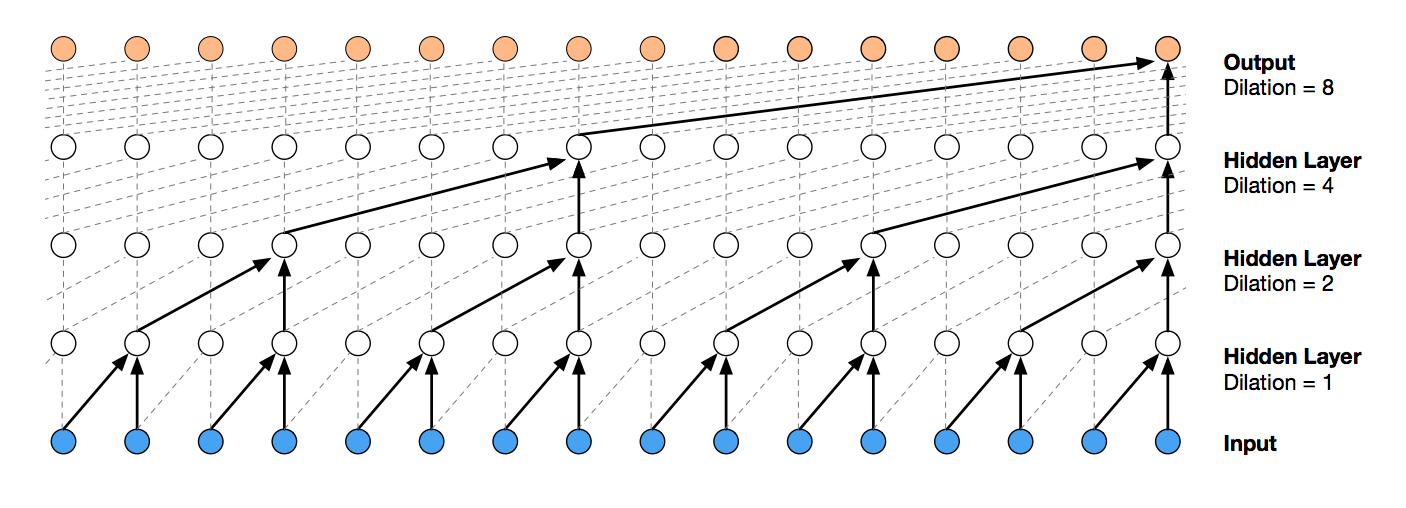}
\caption{A stack of dilated causal convolutions as used by WaveNet, reproduced from~\cite{wavenet}.}
\label{fig:wavenet}
\end{figure}

The modeling of conditional probabilities in WaveNet utilizes causal convolutions, similar to masked convolutions used in PixelRNN and similar image generation networks~\cite{pixrnn}. Causal convolutions ensure that the prediction for time step $t$ only depends on the predictions for previous timesteps. Furthermore, the causal convolutions are dilated; these are convolutions where the filter is applied over an area larger than its length by skipping particular input values, as shown in Figure~\ref{fig:wavenet}. In addition to dilated causal convolutions, each layer features gated activation units and residual connections, as well as skip connections to the final output layers.

\subsection{Symbolic Audio Models}
Most deep learning approaches for automatic music generation are based on symbolic representations of the music. MIDI (Musical Instrument Digital Interface),\footnote{https://www.midi.org/specifications} for example, is a ubiquitous standard for file format and protocol specification for symbolic representation and transmission.  Other representations that have been utilized include the piano roll representation~\cite{pianoroll}---inspired by player piano music rolls---text representations (e.g., ABC notation\footnote{http://abcnotation.com}), chord representations (e.g., Chord2Vec~\cite{chord2vec}), and lead sheet representations. A typical scenario for producing music in such models is to train and generate on the same type of representation; for instance, one may train on a set of MIDI files that encode melodies, and then generate new MIDI melodies from the learned model. These models attempt to capture the aspect of long-range dependency in music.

A traditional approach to learning temporal dependencies in data is to use recurrent neural networks (RNNs). A recurrent neural network receives a timestep of a series $x_t$ along with a hidden state $h_t$ as input. It outputs $y_{t}$, the model output at that timestep, and computes $h_{t+1}$, the hidden state at the next timestep. RNNs take advantage of this hidden state to store some information from the previous timesteps. In practice, vanilla RNNs do not perform well when training sequences have long temporal dependencies due to issues of vanishing/exploding gradients~\cite{rnn_gradient}. This is especially true for music, as properties such as key signature and time signature may be constant throughout a composition.

Long Short Term Memory networks are a variant of RNNs that have proven useful in symbolic music generation systems. LSTM networks modify the way memory information is stored in RNNs by introducing another unit to the original RNN network: the cell state, $c_t$, where the flow of information is controlled by various gates. LSTMs are designed such that the interaction between the cell state and the hidden state prevents the issue of vanishing/exploding gradients~\cite{hochreiter,deepbook}.

There are numerous existing deep learning symbolic music generation approaches~\cite{music_survey}, including models that are based on RNNs, many of which use an LSTM as a key component of the model. Some notable examples include DeepBach~\cite{deepbach}, the CONCERT system~\cite{concert}, the Celtic Melody Generation system~\cite{celtic} and the Biaxial LSTM model~\cite{biaxial}. Additionally, some approaches combine RNNs with restricted Boltzmann machines~\cite{rnnrbm1,rnnrbm2,rnnrbm3,rnnrbm4}.

\section{Architecture}
\vspace{-.2cm}
We first discuss our symbolic method for generating unique melodies, then detail the modifications to the raw audio model for compatibility with these generations. Modifying the architecture involves working with both symbolic and raw audio data in harmony.

\subsection{Unique Symbolic Melody Generation with LSTM Networks}

\begin{figure}[t!]
\centering
\includegraphics[width=1 \columnwidth]{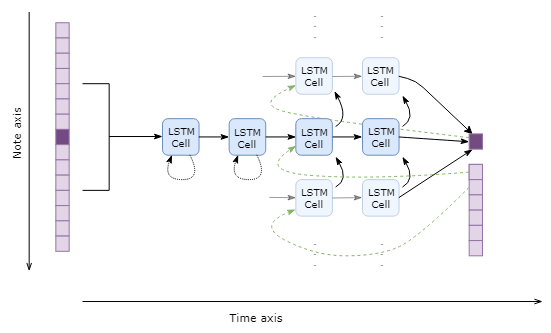}
\caption{A representation of a biaxial LSTM network. Note that the first two layers have connections across timesteps, while the last two layers have recurrent connections across notes~\cite{biaxial}.}
\label{fig:biaxial}
\end{figure}

Recently, applications of LSTMs specific to music generation, such as the biaxial LSTM, have been implemented and explored. This model utilizes a pair of tied, parallel networks to impose LSTMs both in the temporal dimension and the pitch dimension at each timestep. Each note has its own network instance at each timestep, and receives input of the MIDI note number, pitchclass, beat, and information on surrounding notes and notes at previous timesteps. This information first passes through two layers with connections across timesteps, and then two layers with connections across notes, detailed in Figure~\ref{fig:biaxial}. This combination of note dependency and temporal dependency allow the model to not only learn the overall instrumental and temporal structure of the music, but also capture the interdependence of the notes being played at any given timestep~\cite{biaxial}. 

We explore the sequential combination of the symbolic and raw audio models to produce structured raw audio output. We train a biaxial LSTM model on the MIDI files of a particular genre of music as training data, and then feed the MIDI generations from this trained model into the raw audio generator model.

\subsection{Local Conditioning with Raw Audio Models}
\begin{figure}[t!]
\centering
\includegraphics[width=1 \columnwidth]{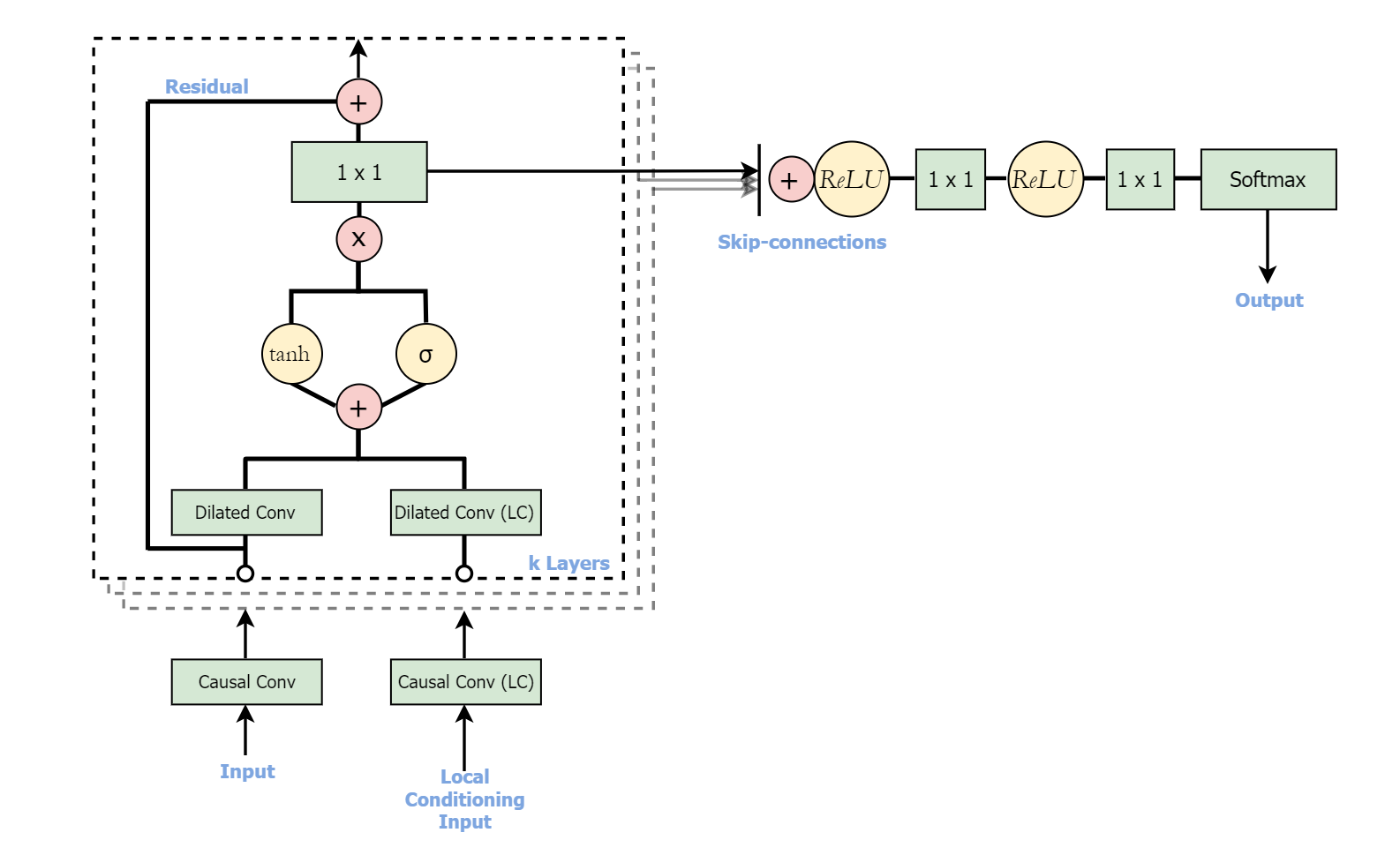}
\caption{An overview of the model architecture, showing the local conditioning time series as an extra input.}
\label{fig:lcwavenet}
\end{figure}

Once a learned symbolic melody is obtained, we treat it as a second time series within our raw audio model (analogous to using a second time series with a desired text to be spoken in the speech domain). In particular, in the WaveNet model, each layer features a gated activation unit.  If ${x}$ is the raw audio input vector, then at each layer $k$, it passes through the following gated activation unit:
\vspace{-.2cm}
\begin{equation}\label{GAU}
{z} = \mbox{tanh}(W_{f,k} \ast {x}) \odot \sigma(W_{g,k} \ast {x}),
\end{equation}
where $\ast$ is a convolution operator, $\odot$ is an elementwise multiplication operator, $\sigma(\cdot)$ is the sigmoid function, and the $W_{f, k}$ and $W_{g,k}$ are learnable convolution filters.  Following WaveNet's use of local conditioning, we can introduce a second time series ${y}$ (in this case from the LSTM model, to capture the long-term melody), and instead utilize the following activation, effectively incorporating ${y}$ as an extra input:
\vspace{-.2cm}
\begin{equation}\label{modifiedGAU}
{z} = \mbox{tanh}(W_{f,k} \ast {x} + V_{f,k} \ast {y}) \odot \sigma(W_{g,k} \ast {x} + V_{g,k} \ast {y}),
\end{equation}
where $V$ are learnable linear projections. By conditioning on an extra time series input, we effectively guide the raw audio generations to require certain characteristics; ${y}$ influences the output at all timestamps.

In our modified WaveNet model, the second time series ${y}$ is the upsampled MIDI embedding of the local conditioning time series. In particular, local conditioning (LC) embeddings are 128-dimensional binary vectors, where ones correspond to note indices that are being played at the current timestep. As with the audio time series, the LC embeddings first go through a layer of causal convolutions to reduce the number of dimensions from 128 to 16, which are then used in the dilation layers as the conditioning samples. This reduces the computational requirement for the dilation layers without reducing the note state information, as most of the embeddings are zero for most timestamps. This process along with the surrounding architecture is shown in Figure~\ref{fig:lcwavenet}.

\subsection{Hyperparameter Tuning}
Table~\ref{wavenet_hyper} enumerates the hyperparameters used in the WaveNet-based conditioned model to obtain our results. We note that the conditioned model needs only 30 dilation layers as compared to the 50 we had used in the unconditioned network. Training with these parameters gave us comparable results as compared to the unconditioned model in terms of the timbre of instruments and other nuances in generations. This indicates that the decrease in parameters is offset by the extra information provided by the conditioning time series.

\section{Empirical Evaluation}
\vspace{-.2cm}
Example results of generations from our models are posted on our web page.\footnote{http://people.bu.edu/bkulis/projects/music/index.html}

One of the most challenging tasks in automated music generation is evaluating the resulting music. Any generated piece of music can generally only be subjectively evaluated by human listeners. Here, we qualitatively evaluate our results to the best of our ability, but leave the results on our web page for the reader to subjectively evaluate. We additionally quantify our results by comparing the resulting loss functions of the unconditioned and conditioned raw audio models. Then, we evaluate the structural component by computing the cross-correlation between the spectrogram of the generated raw audio and conditioning input.

\subsection{Training Datasets and Loss Analysis}

\begin{table}[!t]
	\centering
	\begin{tabular}{| c | c | c |}
    	\hline
    	\bf Instrument & \bf Minutes & \bf Labels\\
    	\hline
    	Piano & 1,346 & 794,532\\
    	Violin & 874 & 230,484\\
    	Cello & 621 & 99,407\\
        Solo Piano & 917 & 576,471\\
        Solo Violin & 30 & 8,837\\
        Solo Cello & 49	& 10,876\\
    	\hline
  \end{tabular}
  \caption{Statistics of the MusicNet dataset. ~\cite{musicnet}}
  \label{musicnet_stats}
\end{table}

\begin{table}[!t]
	\centering
	\begin{tabular}{| l | r |}
    	\hline
    	\bf Hyperparameter & \bf Value\\
    	\hline
        Initial Filter Width & 32\\
        Dilation Filter Width & 2\\
    	Dilation Layers & 30\\
    	Residual Channels & 32\\
        Dilation Channels & 32\\
    	Skip Channels & 512\\
        Initial LC Channels & 128\\
        Dilation LC Channels & 16\\
        Quantization Channels & 128\\
    	\hline
  \end{tabular}
  \caption{WaveNet hyperparameters used for training of the conditioned network.}
  \label{wavenet_hyper}
\end{table}

At training time, in addition to raw training audio, we must also incorporate its underlying symbolic melody, perfectly aligned with the raw audio at each timestep. The problem of melody extraction in raw audio is still an active area of research; due to a general lack of such annotated music, we have experimented with multiple datasets.


Primarily, we have been exploring use of the recently-released MusicNet database for training~\cite{musicnet}, as this data features both raw audio as well as melodic annotations. Other metadata is also included, such as the composer of the piece, the instrument with which the composition is played, and each note's position in the metrical structure of the composition. The music is separated by genre; there are over 900 minutes of solo piano alone, which has proven to be very useful in training on only one instrument. The different genres provide many different options for training. Table~\ref{musicnet_stats} shows some other statistics of the MusicNet dataset.

\begin{figure}[t!]
\includegraphics[width=1\columnwidth]{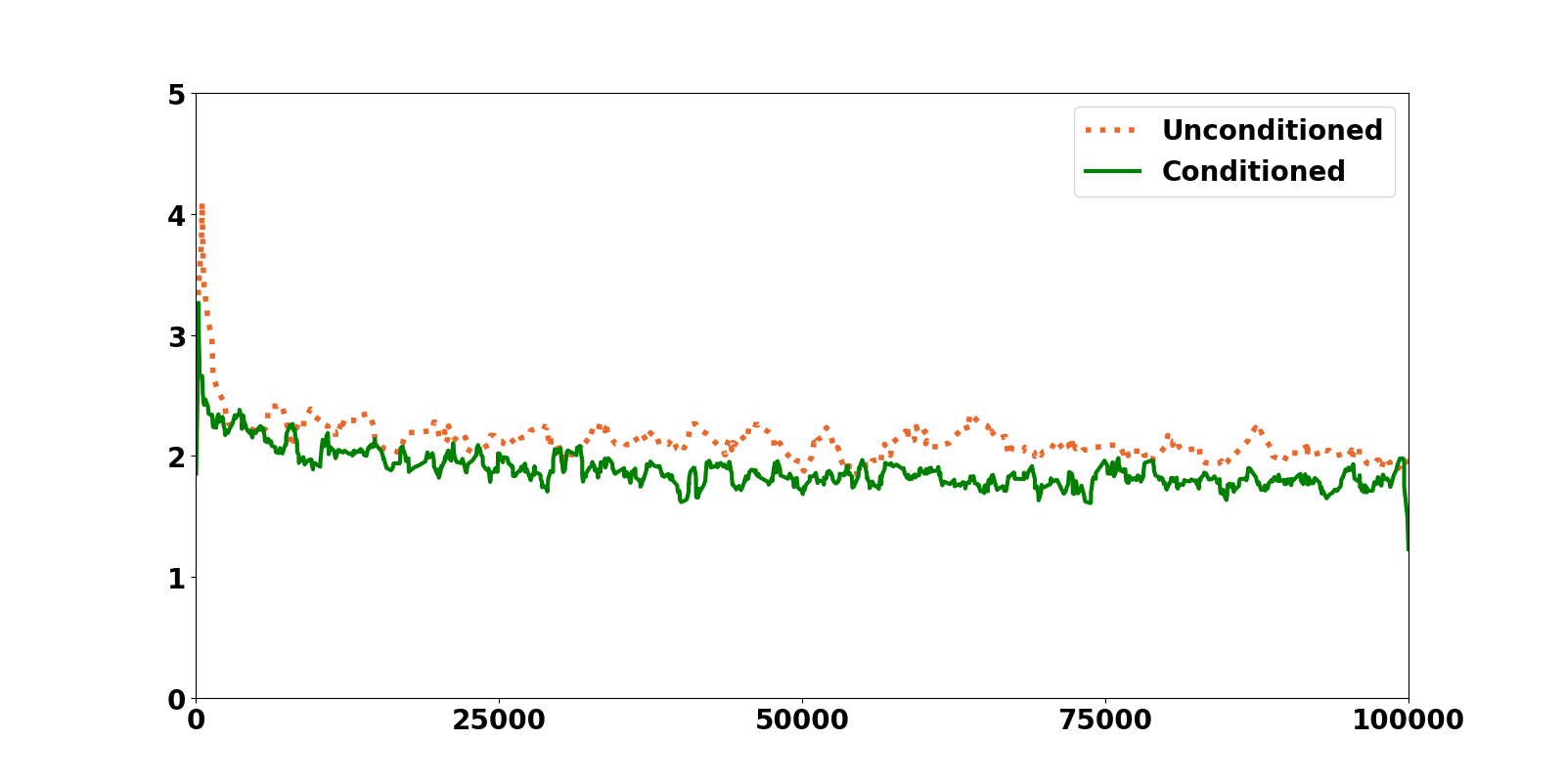}
\caption{Cross entropy loss for the conditioned (solid green) and unconditioned (dotted orange) WaveNet models over the first 100,000 training iterations, illustrating the lower training loss of the conditioned model.}
\label{fig:lossfunc}
\end{figure}

After training with these datasets, we have found that the loss for the unconditioned and conditioned WaveNet models follows our expectation of the conditioned model exhibiting a lower cross-entropy training loss than the unconditioned model. This is due to the additional embedding information provided along with the audio in the conditioned case. Figure~\ref{fig:lossfunc} shows the loss for two WaveNet models trained on the MusicNet cello dataset over 100,000 iterations, illustrating this decreased loss for the conditioned model.

\subsection{Unconditioned Music Generation with WaveNet}

We preface the evaluation of our musical results by acknowledging the fact that we first tuned WaveNet for unstructured music generation, as most applications of WaveNet have explored speech applications. Here we worked in the unconditioned case, i.e., no second time series was input to the network. We tuned the model to generate music trained on solo piano inputs (about 50 minutes of the Chopin nocturnes, from the YouTube-8M dataset~\cite{8m}), as well as 350 songs of various genres of electronic dance music, obtained from No Copyright Sounds\footnote{https://www.youtube.com/user/NoCopyrightSounds}.
\begin{figure}[t!]
\includegraphics[width = \columnwidth]{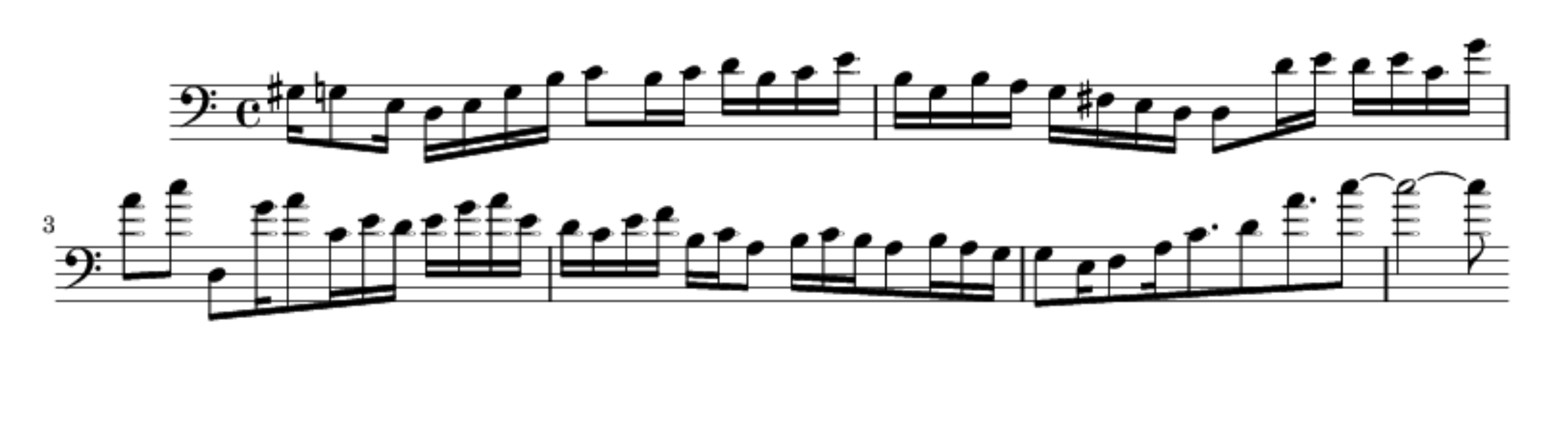}
\caption{Example MIDI generation from the biaxial LSTM trained on cello music, visualized as sheet music.}
\label{fig:lstm_output}
\end{figure}

We found that WaveNet models are capable of producing lengthy, complex musical generations without losing instrumental quality for solo instrumental training data. The network is able to learn short-range dependencies, including hammer action and simple chords. Although generations may have a consistent energy, they are unstructured and do not contain any long-range temporal dependencies. Results that showcase these techniques and attributes are available on our webpage.



\subsection{Structure in Raw Audio Generations}
We evaluate the structuring ability of our conditioned raw audio model for a generation based on how closely it follows the conditioning signal it was given, first using popular existing melodies, then the unique LSTM generations. We use cross-correlation as a quantitative evaluation method. We also acknowledge the applications of our model to edit existing raw audio.

\subsubsection{Raw Audio from Existing Melodies}
We evaluate our approach first by generating raw audio from popular existing melodies, by giving our conditioned model a second time series input of the Happy Birthday melody and a C major scale. Since we are familiar with these melodies, they are easier to evaluate by ear.

Initial versions of the model evaluated in this way were trained on the MusicNet cello dataset. The generated raw audio follows the conditioning input, the recognizable Happy Birthday melody and C major scale, in a cello timbre. The results of these generations are uploaded on our webpage.

\begin{figure}[t!]
\centering
\begin{subfigure}[b]{0.55\textwidth}
   \includegraphics[width=0.9\linewidth]{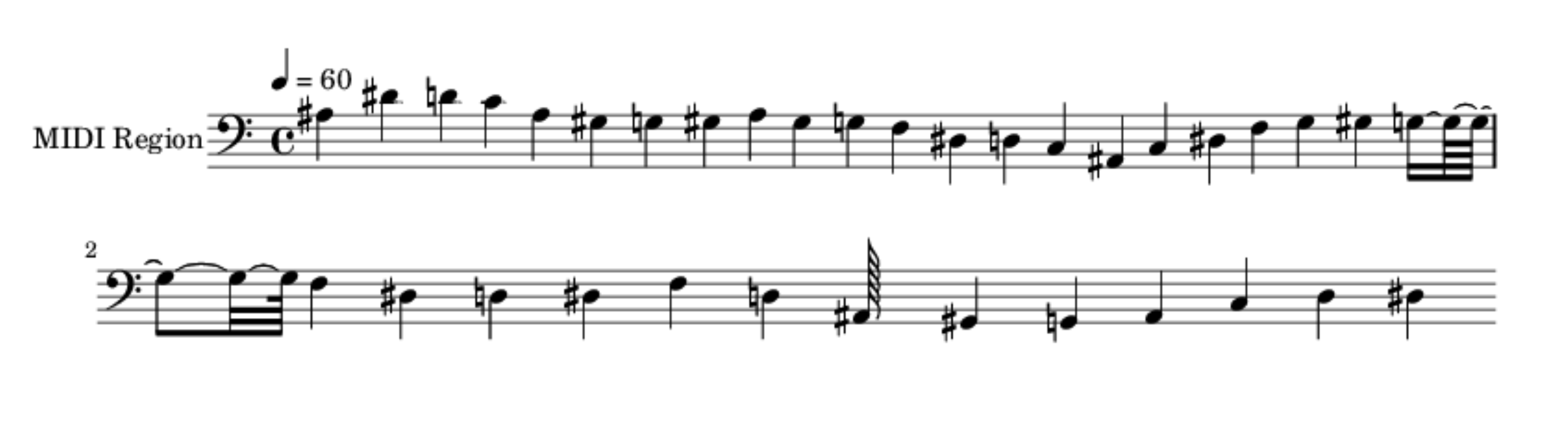}
   \caption{Unedited training sample from the MusicNet dataset.}
   \label{fig:unedited} 
\end{subfigure}

\begin{subfigure}[b]{0.55\textwidth}
   \includegraphics[width=0.9\linewidth]{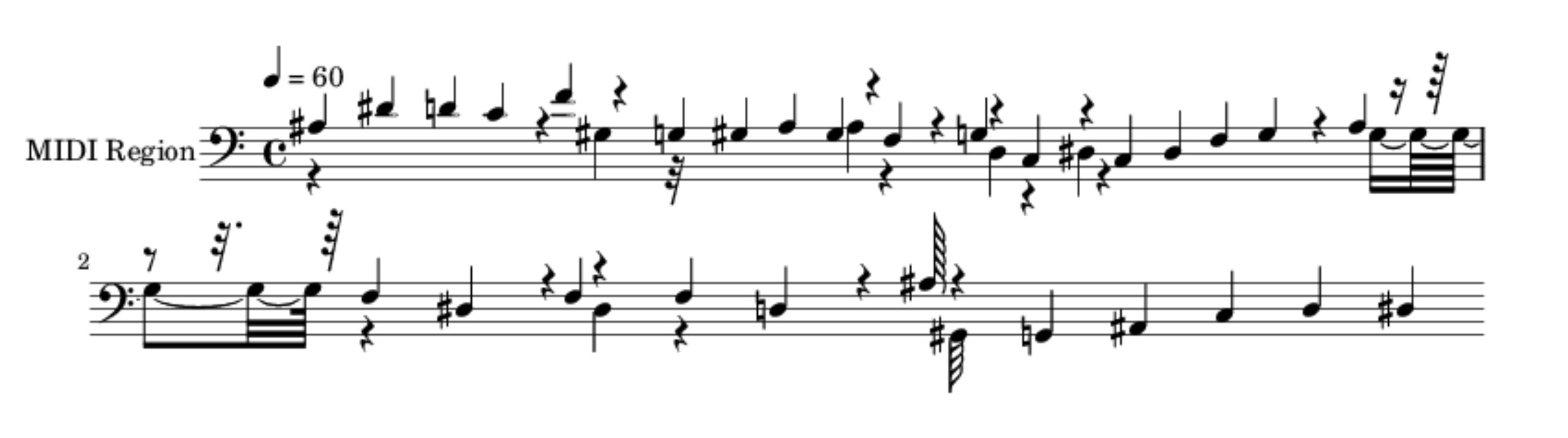}
   \caption{Slightly modified training sample.}
   \label{fig:edited}
\end{subfigure}

\caption{MIDI representations of a sample from the MusicNet solo cello dataset, visualized as sheet music; (b) is a slightly modified version of (a), the original training sample. We use these samples to showcase the ability of our model to ``edit" raw audio.}
\label{fig:edit}
\end{figure}

\begin{figure*}[ht!]
\includegraphics[width = \textwidth]{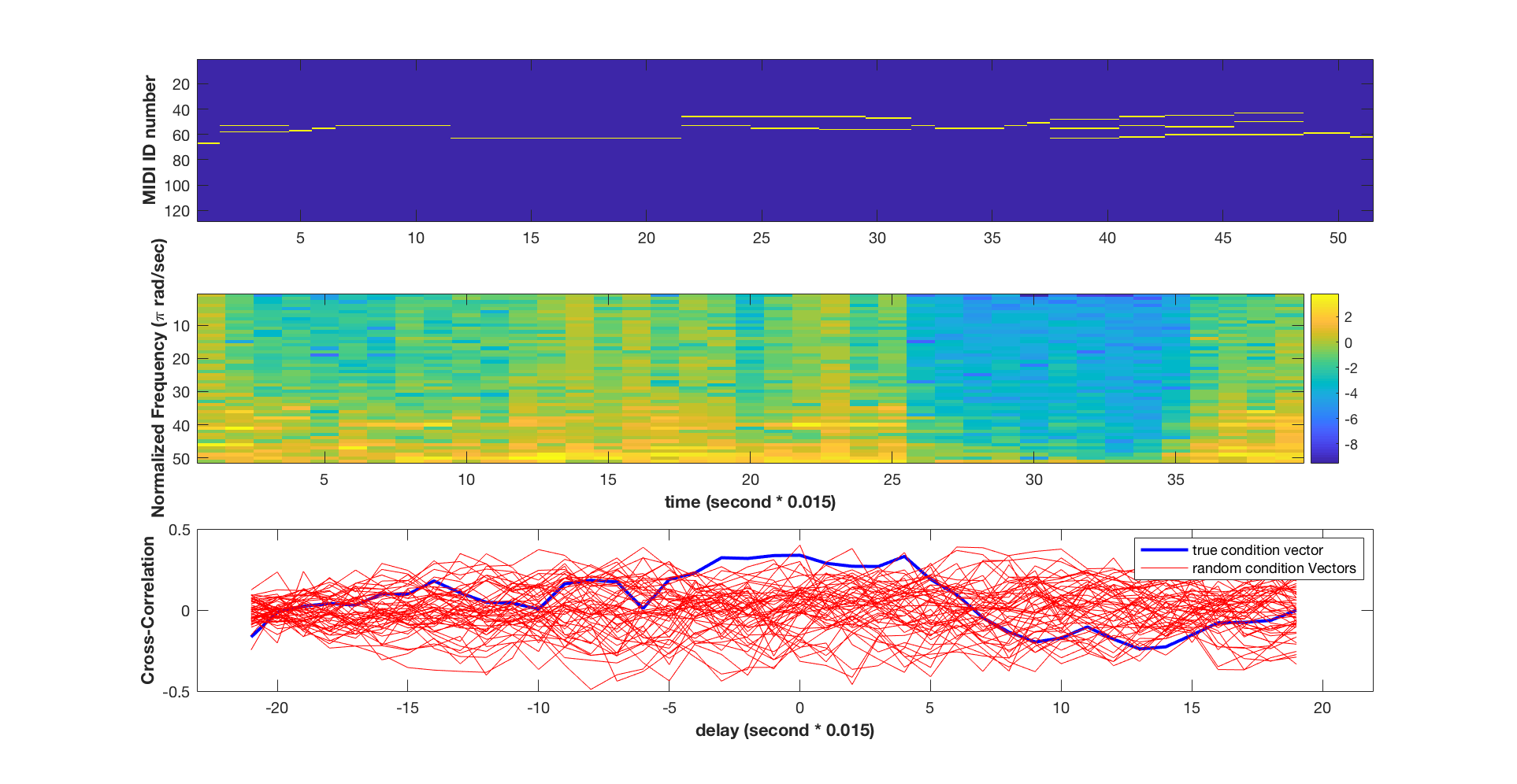}
\caption{Comparison of the novel LSTM-generated melody (top) and the corresponding raw audio output of the conditioned model represented as a spectrogram (middle). The bottom plot shows the cross-correlation between the frequency of the highest note of the MIDI and the most active frequency of raw audio from the WaveNet-based model, showing strong conditioning from the MIDI on the generated audio.}
\label{fig:lstmcor}
\end{figure*}

\subsubsection{Raw Audio From Unique LSTM Generations}
After generating novel melodies from the LSTM, we produced corresponding output from our conditioned model. Since it is difficult to qualitatively evaluate such melodies by ear due to unfamiliarity with the melody, we are interested in evaluating how accurately the conditioned model follows a novel melody quantitatively. We evaluate our results by computing the cross-correlation between the MIDI sequence and the spectrogram of the generated raw audio as shown in Figure~\ref{fig:lstmcor}. Due to the sparsity of both the spectrogram and the MIDI file in the frequency dimension, we decided to calculate the cross-correlation between one-dimensional representations of the two time series. We chose the frequency of the highest note in the MIDI at each timestep as its one-dimensional representation. In the case of the raw audio, we chose the most active frequency in its spectrogram at each timestep. We acknowledge some weakness in this approach, since some information is lost by reducing the dimensionality of both time series.

Cross-correlation is the ``sliding dot product'' of two time series --- a measure of linear similarity as a function of the displacement of one series relative to the other. In this instance, the cross-correlation between the MIDI sequence and the corresponding raw audio peaks at delay 0 and is equal to 0.3. In order to assure that this correlation is not due to chance, we have additionally calculated the cross-correlation between the generated raw audio and 50 different MIDI sequences in the same dataset. In Figure~\ref{fig:lstmcor}, we can see that the cross-correlation curve stays above the other random correlation curves in the the area around delay 0. This shows that the correlation found is not by chance, and the raw audio output follows the conditioning vector appropriately.

This analysis generalizes to any piece generated with our model; we have successfully been able to transform an unstructured model with little long-range dependency to one with generations that exhibit certain characteristics.

\subsubsection{Editing Existing Raw Audio}
\vspace{-1mm}
In addition, we explored the possibility of using our approach as a tool similar to a MIDI synthesizer, where we first generate from an existing piece of a symbolic melody, in this case, from the training data. Then, we generate new audio by making small changes to the MIDI, and evaluate how the edits reflect in the generated audio. We experiment with this with the goal of achieving a higher level of fidelity to the audio itself rather using a synthesizer to replay the MIDI as audio, as that often forgoes the nuances associated with raw audio.

Figure~\ref{fig:edit}(a) and \ref{fig:edit}(b) respectively show a snippet of the training data taken from the MusicNet cello dataset and the small perturbations made to it, which were used to evaluate this approach. The results posted on our webpage show that the generated raw audio retains similar characteristics between the original and the edited melody, while also incorporating the changes to the MIDI in an expressive way.
\vspace{-0.5cm}
\section{Conclusions and Future Work}
\vspace{-.2cm}
In conclusion, we focus on combining raw and symbolic audio models for the improvement of automatic music generation. Combining two prevalent models allows us to take advantage of both of their features; in the case of raw audio models, this is the realistic sound and feel of the music, and in the case of symbolic models, it is the complexity, structure, and long-range dependency of the generations. 

Before continuing to improve our work, we first plan to more thoroughly evaluate our current model using ratings of human listeners. We will use crowdsourced evaluation techniques (specifically, Amazon Mechanical Turk\footnote{https://www.mturk.com/mturk/}) to compare our outputs with other systems. 


A future modification of our approach is to merge the LSTM and WaveNet models to a coupled architecture. This joint model would eliminate the need to synthesize MIDI files, as well as the need for MIDI labels aligned with raw audio data. In essence, this adjustment would create a true end-to-end automatic music generation model.

Additionally, DeepMind recently updated the WaveNet model to improve generation speed by 1000 times over the previous model, at 16 bits per sample and a sampling rate of 24kHz~\cite{newnet}. We hope to investigate this new model to develop real-time generation of novel, structured music, which has many significant implications.

The potential results of our work could augment and inspire many future applications. The combination of our model with multiple audio domains could be implemented; this could involve the integration of speech audio with music to produce lyrics sung in tune with our realistic melody.

Even without the additional improvements considered above, the architecture proposed in this paper allows for a modular approach to automated music generation. Multiple different instances of our conditioned model can be trained on different genres of music, and generate based on a single local conditioning series in parallel. As a result, the same melody can be reproduced in different genres or instruments, strung together to create effects such as a quartet or a band. The key application here is that this type of synchronized effect can be achieved without awareness of the other networks, avoiding model interdependence.

\section{Acknowledgement}
\vspace{-.2cm}
We would like to acknowledge that this research was supported in part by NSF CAREER Award 1559558.

\bibliography{ISMIR2018}

\end{document}